# Large-Scale DNS of Gas-Solid Flow on Mole-8.5

Qingang Xiong[a,b], Guofeng Zhou[a,b], Bo Li[a,b], Ji Xu[a,b], Xiaojian Fang[a,b],

Junwu Wang[a], Xianfeng He[a], Xiaowei Wang[a*], Limin Wang[a*], Wei Ge[a] and Jinghai Li[a]

[a]State Key Laboratory of Multiphase Complex Systems, Institute of Process Engineering (IPE),

Chinese Academy of Sciences (CAS), P. O. Box 353, Beijing 100190, China

[b]Graduate School of Chinese Academy of Sciences, Beijing 100049, China

*Corresponding authors. Email addresses: xwwang@home.ipe.ac.cn, lmwang@home.ipe.ac.cn

Tel:+86 10 6255 5245; Fax:+86 10 6255 8065

**Abstract** Direct numerical simulation (DNS) for gas-solid flow is implemented on a multi-scale supercomputing system, Mole-8.5, featuring massive parallel GPU-CPU hybrid computing, for which the lattice Boltzmann method (LBM) is deployed together with the immersed moving boundary (IMB) method and discrete element method (DEM). A two-dimensional suspension with about 1,166,400 75-micron solid particles distributed in an area of 11.5cm×46cm, and a three-dimensional suspension with 129,024 solid particles in a domain of 0.384cm×1.512cm×0.384cm are fully-resolved below particle scale and distinct multi-scale heterogeneity are observed. Almost 20-fold speedup is achieved on one Nvidia C2050 GPU over one core of Intel E5520 CPU in double precision, and nearly ideal scalability is maintained when using up to 672 GPUs. The simulations demonstrate that LB-IMB-DEM modeling with parallel GPU computing may suggest a promising approach for exploring the fundamental mechanisms and constitutive laws of complex gas-solid flow, which are, so far, poorly understood in both experiments and theoretical studies.

**Keywords**: Direct numerical simulation; Gas-solid flow; GPU; Lattice Boltzmann method; Immersed moving boundary; Massive parallel computing

## Introduction

With dramatic development of computing technology and numerical methods, direct numerical simulation (DNS) of complex flows, which resolves the smallest continuum scale in the flow system directly from Navier-Stokes equation and Newton's law of mechanics, is receiving more and more attention in recent years (Taylor, et al., 2007; Yu and Shao, 2010). It provides an unprecedented tool for exploring the fundamental mechanisms in these flows, for which the capability of most experimental and theoretical methods are proven to be very limited so far.

Compared with DNS of single-phase turbulent flow (Moin and Mahesh, 1998), multi-phase flows such as fluid-particle flow (Nguyen and Ladd, 2005), gas-liquid flow (Yu and Fan, 2009), gas-liquid-solid flow (Deen, et al., 2009) is, in general, computationally more extensive, largely due to the challenges in enforcing the no-slip condition at the phase interface. Conventional numerical approaches such as finite difference (FD) and finite volume (FV) methods are, at current status, cumbersome in this aspect. On the other hand, though particle methods (PM) such as smoothed particle hydrodynamics (SPH) (Monaghan, 1992) and macro-scale pseudo-particle modeling (MaPPM) (Ge and Li, 2001; Ge and Li, 2003) are more adaptive and scalable for describing complex interfaces, as demonstrated in the simulations of gas-solid suspension (Tang, et al., 2004; Ma, et al., 2006; Ma, et al., 2009; Xiong, et al.,

2010), high computational cost and relatively low accuracy pose considerable limitation to its application in larger systems.

Recently, Wang, et al. (2010) successfully coupled a time-driven hard-sphere model -- pseudo particle modeling (PPM) (Ge and Li, 2003) and lattice Boltzmann method (LBM) with immersed moving boundary (IMB) (Cook, et al., 2004) to conduct DNS of gas-solid fluidization with more than 1000 particles in two dimensions. The transition from homogeneous to heterogeneous flow structures were observed when solid/gas density ratio is increased from 2.5 to 1500, and each case was run on one CPU core within only one week (Wang, 2010). This demonstrated that LBM-IMB is an efficient tool for gas-solid DNS. However, to understand the multi-scale structure in gas-solid flow, no matter in two- or three-dimensions, massive parallel computing is by all means necessary.

The newly established GPGPU (general-purpose graphic processing unit) supercomputing system *Mole-8.5* at IPE, CAS (cf. http://www.top500.org/list/2010/06/100), optimized for multi-scale discrete modeling, can deliver 2Petaflops peak performance in single precision and 1Petaflops in double precision. Thanks to its inherent parallelism, the LBM-IMB method can take full advantage of this system, and NVIDIA's CUDA® (computed unified device architecture) has provided a convenient programming platform for its high-performance implementation, of which the very first version is reported in this shorter communication.

## Numerical scheme

LBM, since its invention more than two decades ago (McNamara and Zanetti, 1988) has been a widely used simulation tool in fluid mechanics for its ability to accurately recover Navier-Stokes equations at low Mach number. In this study, the often used and easy-to-implement LBGK model is adopted with D2Q9 scheme in two-dimensions and D3Q19 scheme in three-dimensions. The detailed description of these two schemes can be found in Ref. (McNamara and Zanetti, 1988). For the gas-solid interface coupling to realize no-slip condition, the immersed moving boundary condition proposed by Nobel and Torczynski (1998) is borrowed and finally embedded in the LBGK equations as a body force:

$$f_i(\mathbf{x}+\mathbf{c}_i \Delta t, t+\Delta t) = f_i(\mathbf{x},t) - \frac{1}{\tau}(1-\beta(\varepsilon,\tau))(f_i(\mathbf{x},t) - f_i^{eq}(\mathbf{x},t)) + \beta(\varepsilon,\tau)\Omega_i^S + F_i, \quad (1)$$

where $\beta$ is a weighting function, $F_i$ is the external body force and $\Omega_i^S$ is an additional term accounting for the on-slip interface condition, they are further expressed as

$$\beta = \frac{\varepsilon(\tau-0.5)}{(1-\varepsilon)+(\tau-0.5)}, \quad (2)$$

$$\Omega_i^S = f_{-i}(\mathbf{x},t) - f_i(\mathbf{x},t) + f_i^{eq}(\rho,\mathbf{V}_S) - f_{-i}^{eq}(\rho,\mathbf{u})$$

where $\varepsilon$ named a local solid area/volume fraction which is defined as the area/volume covered by solid particles of the referred lattice (Wang, et al., 2010). The total hydrodynamic forces and torque exerted on a solid particle covering *n* lattices can be summed as

$$\mathbf{F}_f = \frac{h^2}{\Delta t}\sum_n \left(\beta_n \sum_{i=1}^{8} \Omega_i^S \mathbf{c}_i\right),$$
$$\mathbf{T}_f = \frac{h^2}{\Delta t}\sum_n (\mathbf{x}_n - \mathbf{x}_c) \times \left(\beta_n \sum_{i=1}^{8} \Omega_i^S \mathbf{c}_i\right),$$
(3)

For more details, the work of Cook et al. (2004) can be referred. The interactions between solid particles are treated as that of DEM (Zhang, 2004) but without friction. That is,

$$F_{n,ij} = -k_n \xi_n \mathbf{n}_{ij} - \eta_n \mathbf{v}_{n,ij},\qquad(4)$$

where $k_n$ is the stiffness coefficient. $\xi_n$, $\mathbf{n}_{ij}$, $\eta_n$ and $\mathbf{v}_{n,ij}$ are defined as

$$\begin{aligned}\xi_n &= (R_i - R_j) - |\mathbf{r}_i - \mathbf{r}_j|\\ \mathbf{n}_{ij} &= \frac{\mathbf{r}_i - \mathbf{r}_j}{|\mathbf{r}_i - \mathbf{r}_j|}\\ \eta_n &= \sqrt{\frac{(\pi^2 + (\ln e)^2)m_i m_j}{k_n(m_i + m_j)}},\\ \mathbf{v}_{n,ij} &= (\mathbf{v}_i - \mathbf{v}_j)\bullet \mathbf{n}_{ij}\mathbf{n}_{ij}\end{aligned}\qquad(5)$$

where $i, j$ denotes the solid particle $i$ and $j$. $R_i$ is the hydrodynamic radius and $m_i$ is mass of solid particle $i$. $\mathbf{r}$ is position and $\mathbf{v}$ is velocity. The restitution coefficient $e$ quantifies the extent of energy dissipation due to collision. This collision mode is similar to hard-disk/sphere collisions but can be easily carried out on GPU.

**Parallel implementation**

In recent years, GPU simulation of large-scale chemical systems has been very popular (Radeke, et al., 2010), which puts our physical understanding of such systems much more deeply. For large-scale DNS of particle-fluid systems, the difference in particle positions and velocities may reach $10^7$, which requires double precision computing for higher accuracy. Tesla C2050 GPU provides significantly higher double precision performance as compared to its previous GPU products, making this implementation more profitable. Implementation of LBM for pure gas on single GPU is referred to Tolke (2008) and Tolke and Krafczyk (2008), where excellent performance in terms of memory bandwidth and peak performance were obtained. Shared memory was used to maximize the data-writing efficiency during lattice propagation. For the solid particles, an efficient neighbor-list searching algorithm realized by Xu et al. (2010) for their quasi-real time DEM (discrete element method (Cundall and Strack, 1979)) simulations is adopted, which utilizes more memory space to reduce the computation on particle collision. The neighbor-list of each solid particle is established roughly every 100th time step, reducing the time-consuming neighbor-searching computation considerably.

For the inter-phase coupling, due to the complex topological relations when a lattice lies on the boundary of a solid particle especially for three-dimensional case, accurate calculation of $\gamma$ poses a great challenge to GPU implementation because a large amount of flow controls is needed which could result in obvious drop of performance. In addition, the complicated topology of the boundary lattice has to be updated every step, which is also not suitable to GPU implementation. To solve this problem, a widely used, slightly less accurate but faster method, table-looking, is carried out on GPU. The basic idea is depicted by the two-dimensional case in Fig. 1. For a lattice and a solid particle with a given radius, $\gamma$ can be determined by two factors: the distance between the lattice and the origin of the solid particle $d$, and the orientation of the lattice to the solid particle $\theta$, that is

$$\gamma = \phi(d, \theta). \tag{6}$$

In this way, $\gamma$ is pre-calculated and stored into the constant memory of GPU. During the simulation, as long as $d$ and $\theta$ has been calculated, the corresponding $\gamma$ can be immediately indexed within the constant memory with very low memory access latency to save the time for intensive computation of $\gamma$ at each lattice. The computational process of $\gamma$ in three-dimensional case can be done similarly.

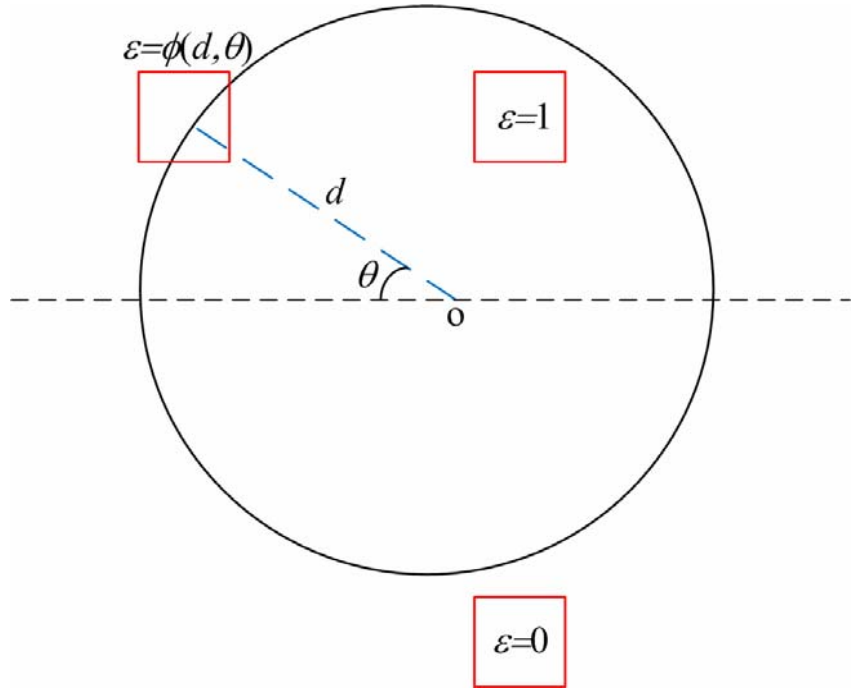

Figure. 1 Determining $\varepsilon$ as a function of lattice position relative to the solid particle.

Due to the memory size limit, only tens million lattices can be calculated on Tesla C2050 and the actual maximum performance is about 200 million lattices update per second (MLUPS) in double precision in three dimensions, which can not meet the demand of practical use where billions of lattices and several millions time steps are needed. So implementation on multi-GPUs distributed in several nodes has to be carried out. In Mole-8.5, six Tesla C2050 GPUs are installed in one node with two quad-core Intel E5520 CPUs and all nodes are connected through high-speed QDR InfiniBand. Each Tesla C2050 GPU can provide 2.5GB global memory and about 1.03 Teraflop computing capability at peak performance. With this hardware architecture, we implement our applications by combining CUDA, shared memory and MPI. However, different from the implementation on single GPU, it brings about extra cost of copying data between CPU and GPU and communication between nodes, which is comparable to the cost of pure GPU kernel executions. For this reason, communications among the CPU cores within a same node are replaced by memory-sharing and MPI is responsible for inter-node data-transfer only. Besides, GPUDirect (Mellanox, 2010), which has been put forward recently, is used to reduce the time for communication among GPUs in different nodes. The flowchart of the hybrid implementation is shown in Fig. 2.

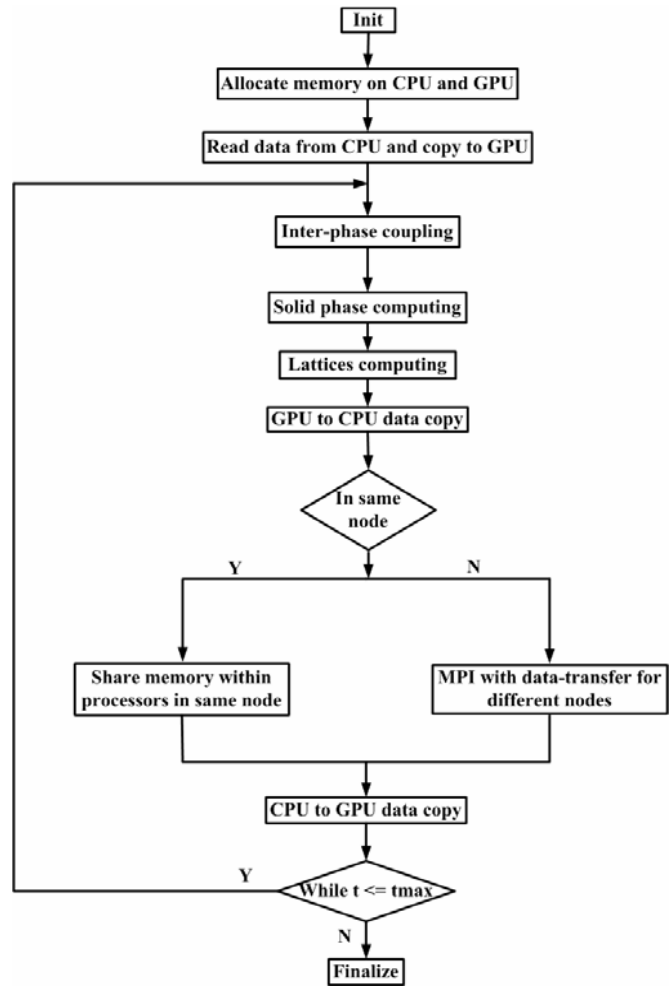

Figure. 2 Flowchart of the CPU-GPU hybrid implementation of gas-solid DNS on Mole-8.5.

## Results and discussions

1. Validation

It has been proved by Strack and Cook (2007) that LBM+IMB is very accurate when particle Reynolds number $Re_p<5$, which covers the entire range of our large-scale DNS. To validate the correctness of our program, three-dimensional gravity-driven flow past a fixed sphere within a cube is performed on a single CPU, multi-GPUs and CPUs with Fluent 6.2®. The physical parameters are shown in Table 1, while the simulation parameters and main results are shown in Table 2. The ratio of particle diameter to lattice spacing ($d/l$) is a compromise between simulation accuracy and computational cost, and Table 1 suggests that $d/l=10$ present a good balance and is hence used hereafter. The difference of $U_g$ between multiple and single GPU implementations is on the order of $10^{-13}$, proving that our parallel GPU program is correct.

| $d/l$ | $U_g$ ($Re_p$=0.2) | Relative error (%) | $U_g$ ($Re_p$=2) | Relative error (%) |
|---|---|---|---|---|
| Fluent 6.2® | 0.05997 | --- | 0.5938 | --- |
| 4 | 0.06956 | 15.99 | 0.7095 | 19.48 |
| 6 | 0.06414 | 6.953 | 0.6724 | 13.24 |

| 8 | 0.06258 | 4.352 | 0.6465 | 8.875 |
|---|---|---|---|---|
| 10 | 0.06169 | 2.868 | 0.6173 | 3.957 |
| 16 | 0.06123 | 2.101 | 0.6104 | 2.796 |
| 20 | 0.06079 | 1.367 | 0.6088 | 2.526 |
| 30 | 0.06036 | 0.6503 | 0.6058 | 2.021 |
| 50 | 0.06017 | 0.3335 | 0.6012 | 1.246 |

Table. 1 Results from the validation case for driven flow past a fixed sphere.

(The relative error is calculated as $| U_g - U_{gF} |/ U_{gF}$.)

2. Simulation results

With the algorithm presented above, a two-dimensional suspension in a domain of 11.5cm×46cm, that is 15360×61440 in reduced values, and a three-dimensional suspension in a domain of 0.384cm×1.512cm×0.384cm, that is, 512×2016×512 in reduced values, are simulated. The physical parameters are also listed in Table 2. Both domains are periodic in all directions, and the numbers of particles embedded are 1,166,400 and 129,024, respectively. These domain sizes and particles numbers are large enough for the suspensions to display continuum properties which can be sampled reasonably and give very useful information to higher level simulation methods such as two fluid model (TFM) (Anderson and Jackson, 1967) and discrete particle model (DPM) (Tsuji, et al., 1993). The solid particles are driven by gravity. A body force $F_i$ is exerted on the gas phase to balance the gravity on solid particles. Initially, for each system, random velocity with the magnitude of $v_{s0}$=5.0×10$^{-3}$ is assigned to each solid particle to destabilize the suspension sooner and hence some computational cost can be saved.

| | Dimensional (2D / 3D) | Dimensionless (2D / 3D) |
|---|---|---|
| Gas density ($\rho_g$) | 1.3 kg/m$^3$ | 1 |
| Solid density ($\rho_s$) | 900 kg/m$^3$ | 692.3 |
| Gas dynamical viscosity ($v$) | 2.308e-5 m$^2$/s | 0.1231 / 0.2462 |
| Body force ($F_i$) | 729.4 m/s$^2$ / 841.7 m/s$^2$ | 8.753e-6 / 4.04e-5 |
| Particle diameter ($d$) | 7.5e-5 m | 10 |
| Gravitational acceleration ($g_s$) | -9.8 m/s$^2$ | -1.176e-7 / 4.704e-7 |
| Particle mass ($m_s$) | 2.982e-11 kg / 1.988 e-10 kg | 54373 / 362491 |
| Restitution coefficient ($e$) | | 0.9 |
| Time step ($\Delta t$) | 3e-7 s / 6e-7 s | 1 |
| Lattice spacing ($l$) | 7.5e-6 m | 1 |
| Stiffness coefficient ($k_n$) | 16 kg/s$^2$ / 27.25 kg/s$^2$ | 2686 / 17892 |
| Terminal velocity ($V_t$) | 0.078624 m/s | 3.14496e-3 / 6.28992e-3 |
| Solid volume fraction ($\varphi$) | | 0.1 / 0.128 |

Table. 2 Parameters for DNS of two- and three-dimensional gas-solid suspensions.

The speedup of a single GPU to one core of a mainstream CPU -- Intel E5520 is measured by modeling three-dimensional suspensions with different domain size. The results are summarized in Table. 3 and we can see that about 20x speedup is obtained. Larger domain size results in slightly higher speedup as it becomes more compute-intensive. The strong scalability of our multi-GPU implementation is analyzed in Figure 3, which remains almost linear below approximately 600 GPUs.

| Domain size (W×H×L) | Steps per second (Fermi GPU) | Steps per second (Intel E5520) | speedup |
|---|---|---|---|
| 32×64×32 | 1189.4 | 65.71 | 18.1 |
| 64×64×64 | 305.7 | 16.44 | 18.6 |
| 64×128×64 | 158.33 | 8.167 | 19.4 |
| 128×128×128 | 40.25 | 2.043 | 19.7 |
| 128×256×128 | 22.23 | 1.056 | 21.0 |

Table. 3 Computation steps per second for a single GPU and a single CPU.

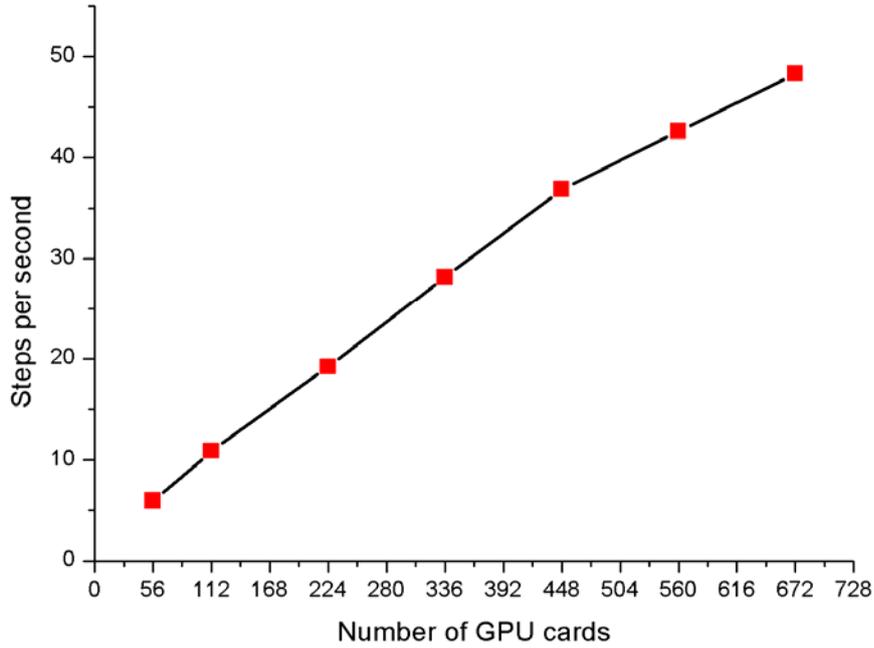

Figure. 3 Scalability of the multi-GPU implementation.

The snapshot of the two-dimensional suspension at four instants is plotted in Figure 4, where distinct particle clusters can be observed. In such large domain, they are fully developed, with a statistically stabilized size distribution and displays kaleidoscopic morphology due to their interactions with surrounding gas flow. The clusters may contain only several particles, but some may have tens of thousands of particles, which has not been seen in DNS previously. The gas velocity inside clusters is significantly lower than that in dilute regions which implies that the drag coefficient of a particle inside clusters is much larger than that of a particle in the dilute phase, supporting the assumptions made in the EMMS (energy minimization multi-scale) model (Li and Kwauk, 1994; Li, et al., 1999; Li, et al., 2005). The temporal evolution of area-averaged inter-phase velocity $V_{ry}$ and drag force $F_{dy}$ is displayed in Figure 5. After time-averaging, the steady-state slip velocity $\overline{V}_w$, 0.017 in this case, is significantly higher than that of a homogeneous suspension predicted by the well-known Wen and Yu equation (Wen and Yu, 1966). This result is even in quantitative agreement to the predictions of the EMMS model (Yang, et al., 2004; Lu, et al., 2009). Three-dimensional DNS gives qualitatively similar conclusions but due to the size limitations, large cluster is still absent and the increase of slip velocity is not so evident, as seen in Figures 6 and 7. Using these large-scale but fully-resolved results, we believe that more meaningful information can be obtained and eventually improve the accuracy of higher level models.

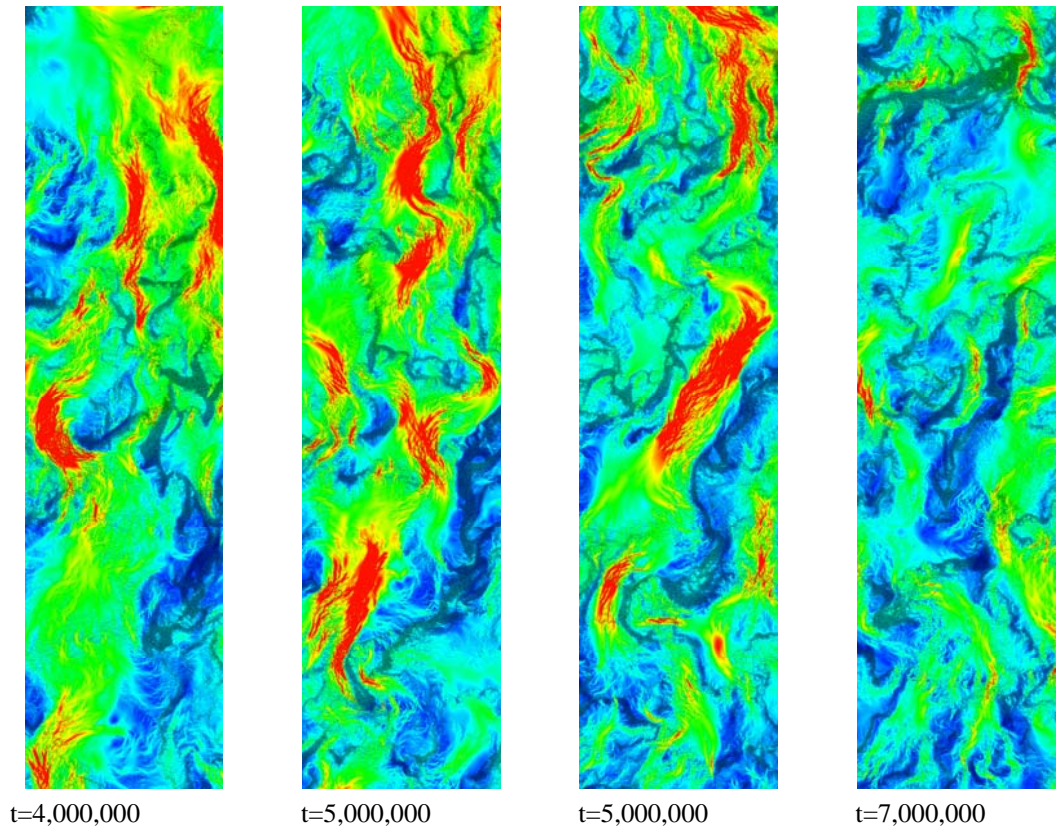

t=4,000,000    t=5,000,000    t=5,000,000    t=7,000,000

Figure. 4 Particle configuration and gas flow field of the whole region in the two-dimensional suspension at four instants.

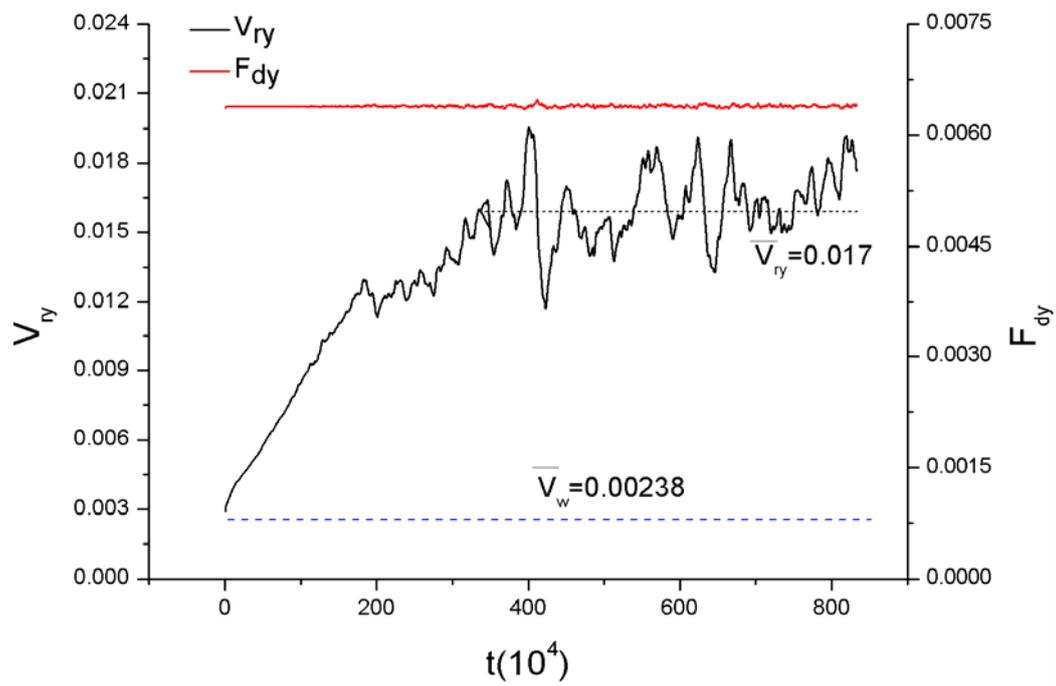

Figure. 5 Temporal evolutions of the area-averaged inter-phase slip velocity and drag force in the two-dimensional suspension.

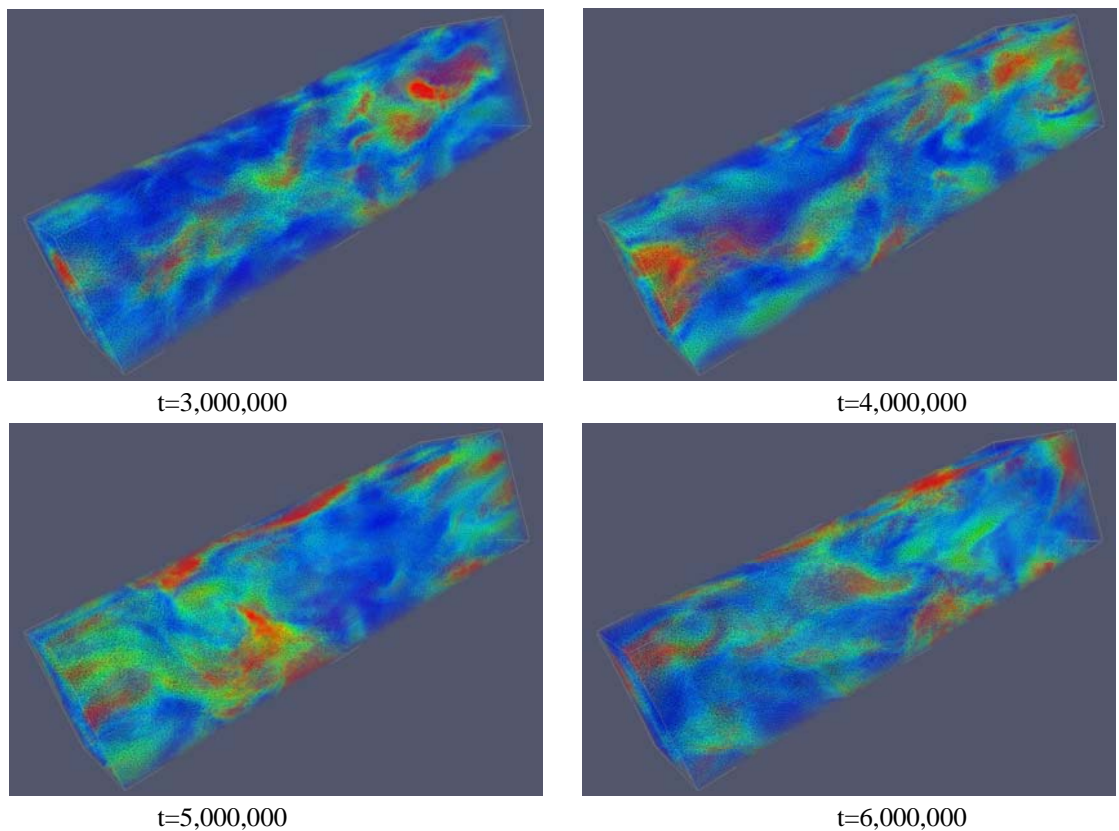

| t=3,000,000 | t=4,000,000 |
| t=5,000,000 | t=6,000,000 |

Figure. 6 Particle configuration and gas flow field of the whole region in the three-dimensional suspension at four instants.

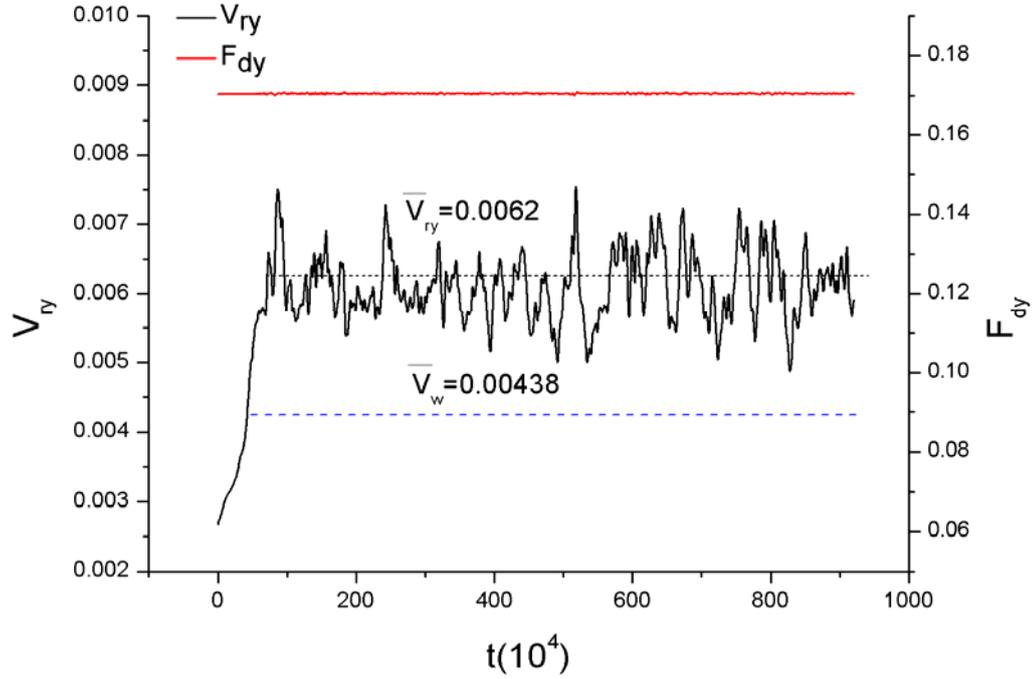

Figure. 7 Temporal evolutions of the area-averaged inter-phase slip velocity and drag force in the three-dimensional suspension.

## Conclusions and prospects

DNS of gas-solid suspensions on a GPGPU cluster through a parallel implementation of LBM coupled with IMB for gas flow and DEM for solid particle motion is presented, where CUDA, shared-memory and MPI programming are combined to achieve higher efficiency. A two-dimensional suspension with 1,166,400 particles and a three-dimensional case with 129,024 particles are simulated and multi-scale clustering of particles is observed, which gives quantitative support to the EMMS model for heterogeneous multi-phase systems. The implementation achieved a speedup of about 20x over one core of a mainstream CPU and good scalability is obtained up to 600 GPUs, demonstrating it as a promising and efficient tool for DNS of gas-solid systems. However, the PCIE data transfer rate is still a limitation in this first version of the parallel implementation, but it can be avoided by using asynchronous communication techniques to overlap computation and data transfer, though much effort is needed for the complex topology in three dimensions.

## Acknowledgement

The authors gratefully acknowledge the financial supports of Ministry of Finance under the Grants nos. ZDYZ2008-2, National Nature Science Foundation of China under the Grants nos. 20221603, 20906091 and 2008BAF33B01, and the Chinese Academy of Sciences under the KGCX2-YW-124. We also thank the group of Zhaoqi Wang from Institute of Compute Technology, CAS, for their collaboration in visualizing our 3D simulation results.